# Germanium-tin (GeSn) avalanche photodiode with up to 2.7 μm cutoff wavelength for extended SWIR detection

*Quang Minh Thai [1], Rajesh Kumar [1], Justin Rudie [1,2], Xiaoxin Wang [3], Abdulla Said Ali [1,2], Perry C. Grant [4], Hryhorii Stanchu [5], Yunsheng Qiu [1], Steven Akwabli [1], Chun-Chieh Chang [6], Jifeng Liu [3], Baohua Li [4], Wei Du [1,2,5] and Shui-Qing Yu [1,2,5]*

[1] Department of Electrical Engineering and Computer Science, University of Arkansas, Fayetteville, Arkansas 72701, USA

[2] Material Science and Engineering Program, University of Arkansas, Fayetteville, Arkansas 72701, USA

[3] Thayer School of Engineering, Dartmouth College, Hanover, New Hampshire 03755, USA

[4] Arktonics, LLC, 1339 S. Pinnacle Dr., Fayetteville, Arkansas 72701, USA

[5] Institute for Nanoscience and Engineering, University of Arkansas, Fayetteville, Arkansas 72701, USA

[6] Center for Integrated Nanotechnologies, Los Alamos National Laboratory, Los Alamos, New Mexico, 87545, USA

**ABSTRACT:** Separate absorption - charge multiplication (SACM) germanium-tin on silicon (GeSn on Si) avalanche photodiode (APD) offers a viable solution to achieve CMOS compatible, high sensitivity detection technology in SWIR/extended SWIR (e-SWIR) range, leveraging the excellent k-factor of Si as multiplication layer and SWIR/e-SWIR band absorption of GeSn. However, unlike well-established growth of GeSn on Si with thick Ge buffer in-between to reduce threading dislocation density due to lattice mismatch, GeSn on Si APD design requires relatively thin Ge buffer to limit electric field drop through the background p-doped buffer and efficiently transporting photocarrier from GeSn absorber to Si multiplication layer, therefore making growth of high Sn content APD for e-SWIR coverage very challenging. In this work, we experimentally demonstrate GeSn on Si APD up to 12.7 % Sn, monolithically grown on Si substrate with 122-nm-thick Ge buffer in between, which is considerably thinner than widely used 700-900 nm thick Ge buffer. Stronger relaxation of GeSn absorber via thin Ge buffer favors Sn incorporation, leading to higher Sn content than the nominal target of 8% Sn. Device detection range is significantly improved compared to previous work - with cutoff wavelength increased up to 2.7 µm at 300 K – in parallel with high avalanche gain at 77 K up to 21 at 1.55 µm and up to 52 at 2 µm, and good responsivity in SWIR/ e-SWIR range, up to 1.45 A.W$^{-1}$ at 1.55 µm and 0.66 A.W$^{-1}$ at 2 µm.

## INTRODUCTION

Photodetectors operating in the short-wave infrared (SWIR) and extended SWIR (e-SWIR) range, with detection range up to 2.5 µm are of particular interest for light imaging, detection and ranging (LiDAR) either in extreme condition of dust, fog and smoke, or in urban area thanks to specific "eye-safe" wavelengths close to 1.55 µm and 2 µm. Avalanche photodiode (APD) is a popular choice for LiDAR application, thanks to its high responsivity resulting from the avalanche multiplication of photocarrier under very high electric field. Beside commercial InGaAs on InP SWIR APD (1.7 µm cutoff wavelength), research efforts continue on other materials system to extend the detection range into e-SWIR and mid-wave infrared (MWIR) range. Steady progress regarding device responsivity and signal-to-noise ratio have been reported for lattice-matched grown APD structure like HgCdTe on CdZnTe substrate [1–10] or As- and Sb- alloys on GaSb/ InAs substrate [11–21], the latter with separate absorber – charge multiplication (SACM) design to freely optimize the absorber thickness and composition, independent of the multiplication region electronic properties. At the same time, significant effort is also directed towards developing Si-based SWIR and e-SWIR APD, to simultaneously benefit from the very low excess noise of Si multiplication region, low cost and scaling production power resulting from its CMOS compatibility, while targeting a comparable performance to CdHgTe or AlInAsSb APD. Several reports on high performance Ge/Si SACM APD in SWIR range have confirmed the feasibility of such approach [22–25]. Recently, successful demonstration of high-quality monolithic growth of germanium-tin (GeSn) alloys on Si - with intermediate Ge buffer layer to accommodate their lattice mismatch - generated a fresh momentum for research on Si-based photodetector in general and APD in particular, with detection range potentially reaching e-SWIR and even MWIR range thanks to the decrease of GeSn band gap with higher Sn content [26–38]. Indeed, GeSn on Si SACM APD with high responsivity at 1.55 µm has been reported, reaching 14.7 A.W$^{-1}$ under 1 µW optical power [36], or with long detection range reaching up to 2.14 µm, as reported in Refs. [38].

Despite these promising results, significant obstacles remain for the development of GeSn on Si APD. First, to extend its effective detection range to e-SWIR and MWIR, a thicker GeSn absorber with higher Sn content above 10% needs to be grown, putting more challenges on the growth process such as low temperature with low growth rate, and crystal quality due to higher lattice mismatch. Second, recent works raised questions on whether the GeSn absorber can be fully depleted for maximum responsivity, between the punch-through and breakdown voltage. This was due to the high background p-doping concentration around $10^{16} - 10^{17}$ cm$^{-3}$ anticipated in both the Ge buffer and GeSn absorber [26,39–41], leading to a rapid electric field drop in both the buffer and absorber, even under high reverse bias. High background p-doping also restricted GeSn growth on very thin Ge buffer between 100 and 300 nm [32–34,37,38] or even direct GeSn growth on Si substrate [35,36], to maintain a sufficient electric field in the absorber for photocarrier transport. As a result, structure design and epitaxial growth of GeSn on Si APD operating in e-SWIR/MWIR range became more challenging, with its feasibility remaining much in doubt.

In this paper, we demonstrated a GeSn on Si APD consisting of 250 nm thick GeSn absorber layer with up to 12.7% Sn content and 122 nm thick Ge absorber. The successfully incorporated high Sn content is attributed to high degree relaxation in the GeSn absorber induced by the thin Ge buffer, which extends detection cutoff wavelength up to 2.7 µm at room temperature, confirming the feasibility of high Sn content GeSn on Si APD in the full e-SWIR range. High avalanche gains at 77 K up to 21 at 1.55 µm and up to 52 at 2 µm were recorded, alongside good responsivities up to 1.45 A.W$^{-1}$ at 1.55 µm and 0.66 A.W$^{-1}$ at 2 µm.

**RESULTS AND DISCUSSIONS**

**Material growth and APD characterization results**

The APD structure was grown using an ASM Epsilon® 2000 Plus reduced pressure chemical vapor deposition (RPCVD) reactor. SiH4, GeH4 and SnCl4 were used as Si, Ge and Sn gas precursors, respectively. The design of GeSn on Si SACM APD structure was shown in **Figure 1a**, with detailed layer thickness and doping concentration in charge and contact layers shown in

**Table 1**. **Figure 1b** showed TEM image of the grown APD structure. A 583 nm thick Si multiplication layer was first grown on n+ Si substrate, with the very top layer (around 100 nm) intended to be p+ doped as charge layer. Ge buffer of 122 nm thick was then grown, followed by the growth of GeSn absorber of 275 nm thick, with the top 25 nm p+ doped for contact layer. For the GeSn absorber, the nominal 8% Sn recipe was applied. Based on our previous study, due to spontaneous-relaxed enhanced (SRE) effect, the Sn content will be gradually increased with film relaxation [42,43]. Therefore, a higher final Sn content was expected. **Figures 1c,d** showed XRD characterization results on the APD structure with 2θ-ω and (-2 -2 4) reciprocal space mapping (RSM) plots: higher Sn content than the nominal value were indeed observed, with three distinct values: 7.8% Sn, 11.6% Sn and 12.7% Sn. The significant increase of Sn content compared to nominal target was probably due to presence of very thin Ge buffer, which amplified the lattice parameter mismatch effect between GeSn layer and Si substrate and introduced additional tensile strain on the Ge buffer itself, in contrast to traditional GeSn growth with very thick, relaxed Ge buffer. SIMS data were plotted in **Figure 1e**, showing Sn content from 7.4% to 13.1%, in line with values extracted from XRD results. It is also worth noting that SIMS data suggested a p+ doping diffusion from the Si charge layer to the Ge buffer, making p+ Ge/Si charge layer instead of pure p+ Si charge layer from the nominal design.

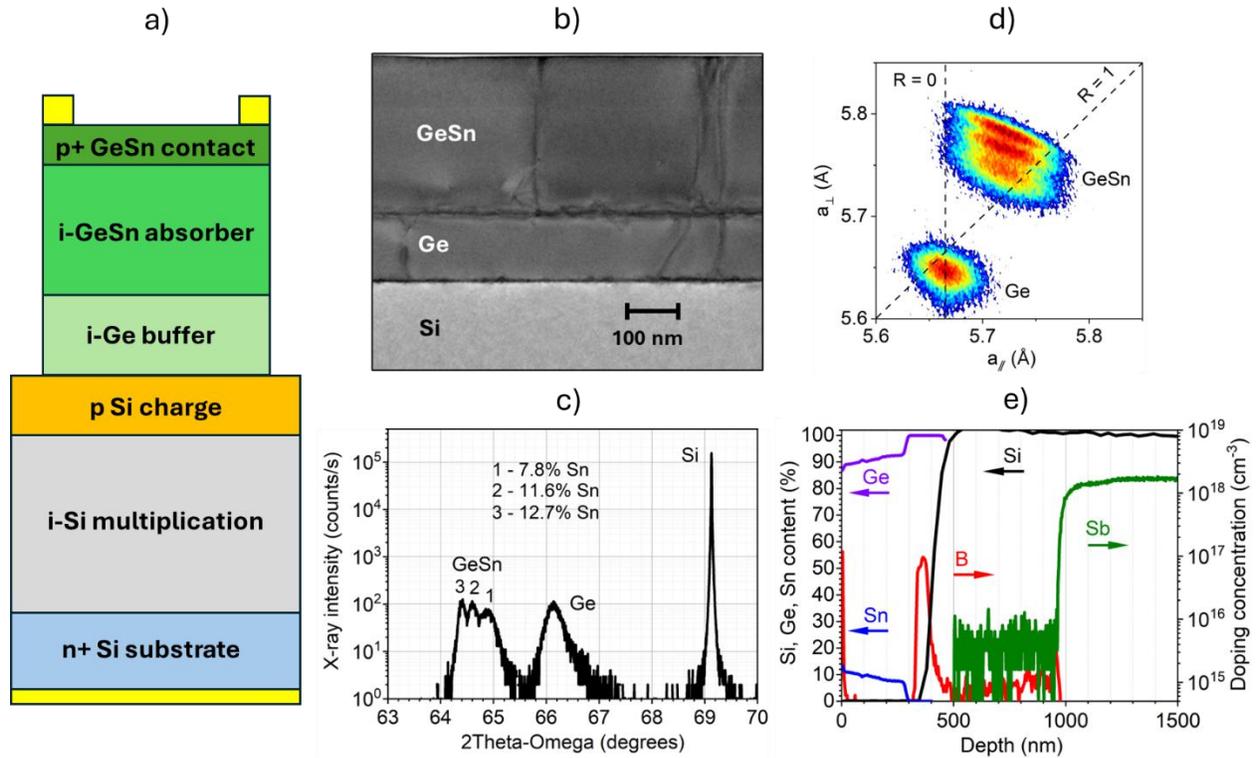

*Figure 1*. a) Nominal design for GeSn on Si APD structure used in this work. b) TEM image of the APD structure, showing clear Ge/Si and GeSn/Ge interfaces. Fading interface lines can also be observed on the GeSn absorber, signaling the split into three layers of different Sn content. c), d) XRD data, with 2θ-ω plot and (-2-2 4) RSM respectively, confirming the presence of three different Sn contents in the GeSn absorber. e) SIMS data, with Si, Ge, Sn content and B concentration (p+ doping). Sb concentration represented n+ doping profile in Si substrate.

*Table 1*. Layer thickness and doping concentration measured for GeSn on Si APD structure in this work.

| Layer | Thickness (nm) | Doping concentration (cm$^{-3}$) |
|---|---|---|
| p+ GeSn contact | 25 | $1.0 \times 10^{17}$ |
| i-GeSn absorber | 250 | |
| i-Ge buffer | 122 | |
| p+ Si/Ge charge | 59 | $1.0 \times 10^{17}$ |
| i-Si multiplication | 583 | |
| n+ Si substrate | | $1.7 \times 10^{18}$ |

Temperature-dependent characterization was performed on a GeSn on Si APD device of 350 µm diameter, with FTIR spectral response and I-V characteristics shown in **Figure 2**. Spectral response revealed a significant improvement in detection range, with detection cutoff wavelength spanning from 2.2 µm (77 K) to 2.7 µm (300 K), showing a clear increase compared to our previous work [38] and to other GeSn on Si SACM APD reported so far (**Figure 2a**). This is indeed attributed to the reduced band gap of GeSn at higher Sn content. Dark IV from 77 K to 300 K were shown in **Figure 2b**, with breakdown voltage can be identified around -21 V at 77 K, 150 K and 200 K through fast increase of dark current, in line with previous results reported for Ge on Si or GeSn on Si APD with Si multiplication region of comparable thickness [25,35–37]. Dark and light IV, with extracted 1.55 µm and 2 µm responsivity showed a rapid increase of responsivity near the breakdown voltage with avalanche gain (**Figures 2c,d**). It is interesting to note that light IV form at low reverse bias (below -12 V) was substantially different between 1.55 µm and 2 µm: weak, but distinguishable photoresponse was observed under 1.55 µm illumination, while no photoresponse was observed under 2 µm illumination in this bias range. In this bias range, we expected that junction width remained low and only expanded to the Ge buffer. In the meantime, due to strong presence of bulk defect from the growth of high Sn content GeSn on thin Ge buffer, photocarrier diffusion length was expected to be low. As a results, at low reverse bias, the majority of photocarrier entering the Si multiplication region were generated in the Ge buffer, creating a weak photoresponse at 1.55 µm and no photoresponse at 2 µm, corresponded to the absorption spectrum of Ge at cryogenic temperature with cutoff wavelength below 1.55 µm. At 77 K, under 100 µW of optical power, 1.55 µm and 2 µm reached up to 1.03 A.W$^{-1}$ and 0.53 A.W$^{-1}$, respectively. Higher responsivity at 1.55 µm was explained by a higher GeSn absorption coefficient at this wavelength compared to 2 µm, as extracted from ellipsometry data (see **Supporting Information, Figures SI1**). Responsivity data from 150 K to 300 K can be found in **Supporting Information, Figure SI2**.

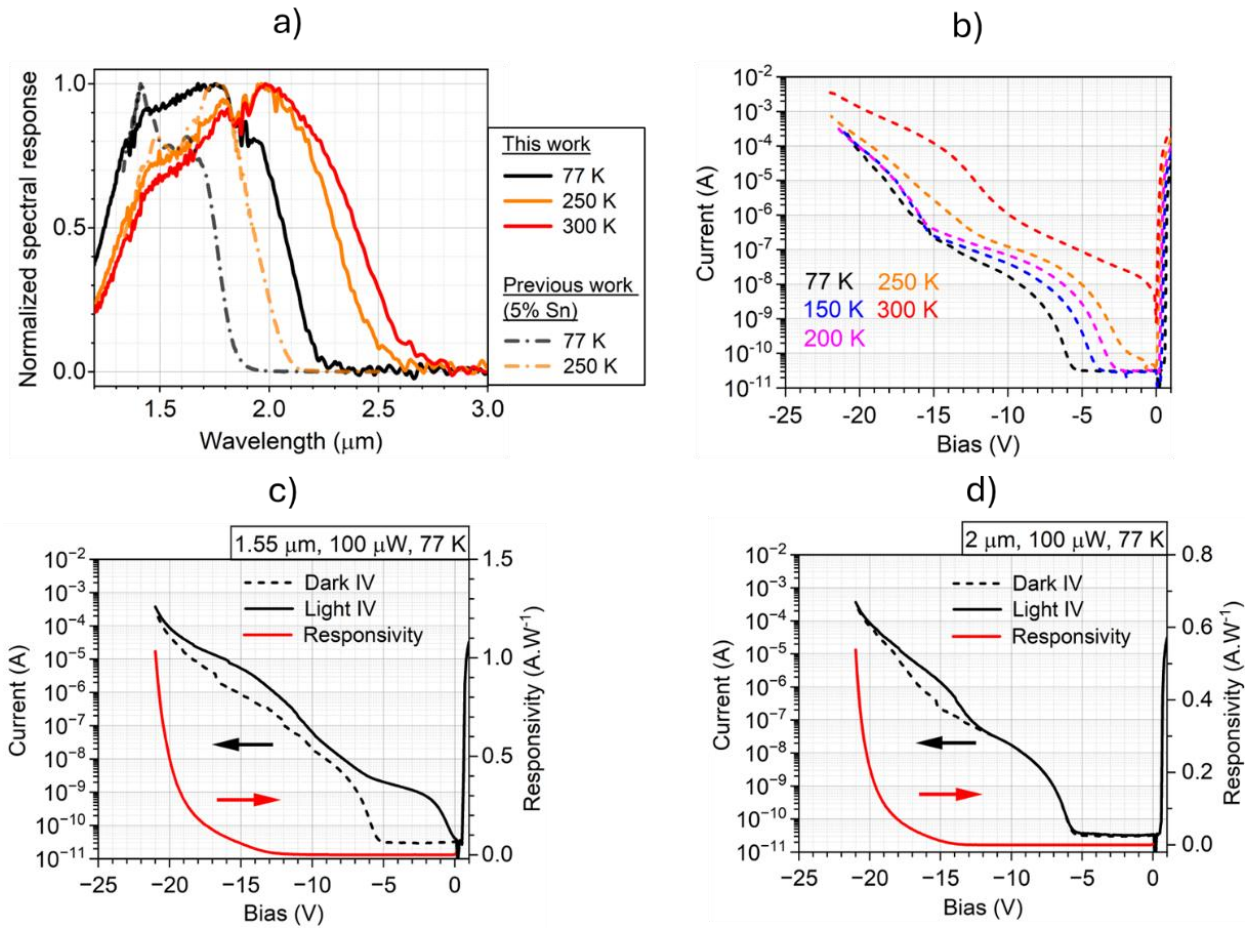

*Figure 2. a) Normalized spectral response of GeSn on Si APD in this work (solid lines) at 77 K, 250 K and 300 K, compared to our previous work [38] (dashed-dot lines). b) Dark I-V characteristics of GeSn on Si APD in this work, from 77 K to 300 K. c), d) Dark and light IV characteristics, with extracted responsivity as function of reverse bias at 77 K under 1.55 µm and 2 µm illumination, respectively. 100 µW of optical power was used in both cases.*

C-V measurement (**Figure 3a**) revealed a dip followed by stabilization of capacitance at high reverse bias – signature of punch-through regime in APD – and showed good agreement with C-V and electric field simulation results (**Figures 3b,c**): for example, at 77 K, simulation predicted a noticeable presence of electric field in GeSn absorber starting from -15 V, well matched to the observed capacitance plateau and thus the punch-through regime. Interestingly, some deviation can be observed between simulation and experiment, where simulation results predicted lower capacitance values overall, in addition to a second capacitance dip at higher bias associated with a continuous increase of the junction width in the GeSn absorber. This is not the case in the experimental results, where capacitance is stable in the punch-through regime, suggesting a fixed junction width value instead, which remained very likely significantly shorter than the GeSn absorber depth. It might be due to stray capacitance resulting from bulk defect (see TEM image in **Figure 1b**) and surface defect (device surface not passivated); detailed analysis of such phenomenon is beyond the scope of this paper. We took the bias at the start of capacitance plateau - which are -15 V at 77 K, -17 V at 150 K, -18 V at 200 K and -19 V at 250 K - as punch-through bias to determine the primary responsivity and thus the avalanche gain of the device (**Figures 3d,e**). At 300 K, capacitance plateau was not observed at high reverse bias, which might signal a punch-through bias beyond our C-V analyzer instrument limit of -20 V; avalanche gain was thus undefined at this temperature. Under 100 µW of optical power, 1.55 µm and 2 µm avalanche gain peaked at 18 and 48 at 77 K, before rapidly decreasing with increased temperature. A significant difference in avalanche gain at 77 K is observed between the two wavelengths, similar to previous report [38]. This is attributed to Franz-Keldysh effect, where optical absorption coefficient can be enhanced near the absorption edge under very strong electric field, as seen at 77 K with detection cutoff wavelength very close to 2 µm (**Figure 2b**). Responsivity and gain decrease with temperature can be attributed to increased loss of electron photocarrier kinetic energy to lattice vibration (phonon), leading to less efficient ionization process. Finally, we measured the dependence of responsivity and gain on the optical power, from 20 µW to 500 µW (**Figure 4**): data showed an increase in responsivity and avalanche gain as optical power decreases, up to 1.45 A.W$^{-1}$ and a gain of 21 under 1.55 µm illumination and up to 0.66 A.W$^{-1}$ and a gain of 52 under 2 µm illumination. It was explained by the charge screening effect from photocarrier: photocarrier-induced electric field drop was mitigated under low optical power, resulting in more efficient avalanche process and thus higher gain and responsivity.

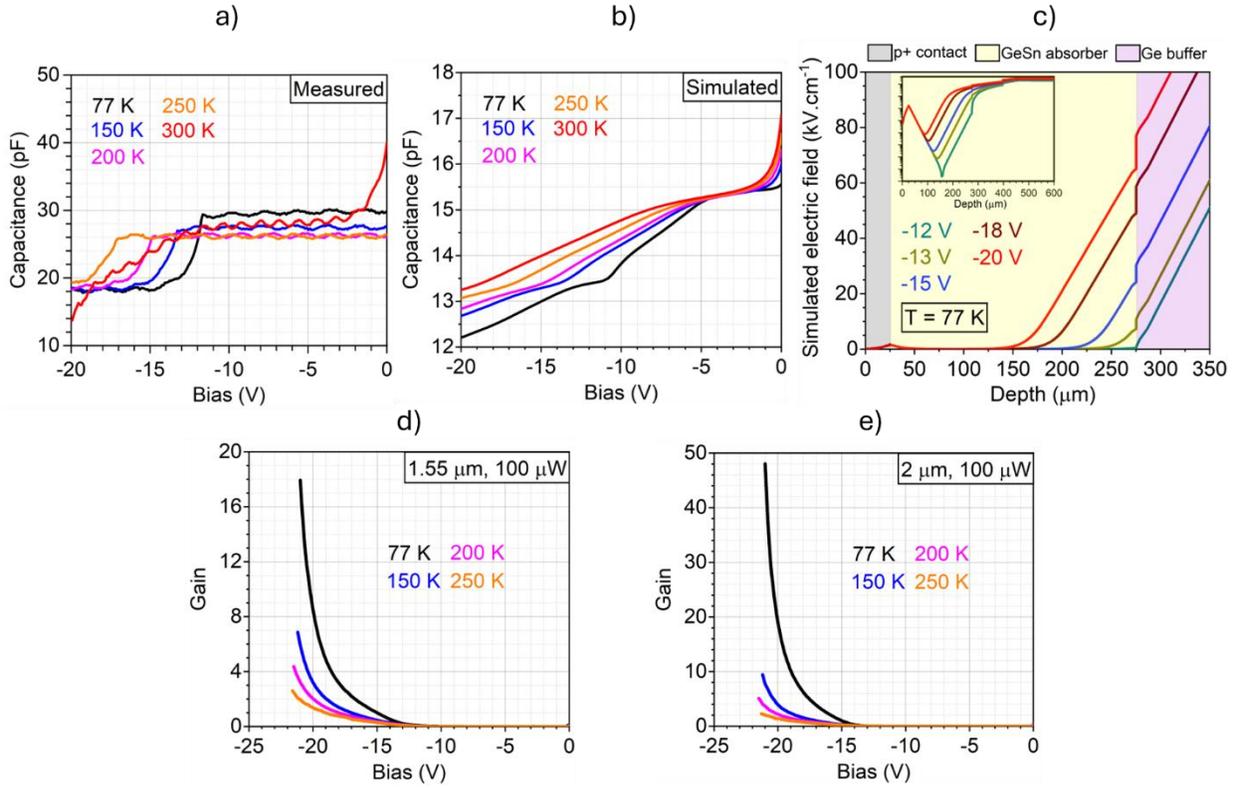

*Figure 3*. a) Measured C-V characteristics on GeSn on Si APD, from 77 K to 300 K. b) Simulated C-V characteristics from 77 K to 300 K. Background p-doping concentration was set at $1 \times 10^{17}$ $cm^{-3}$ in both Ge buffer and GeSn absorber. The charge layer p-doping concentration was allowed to slightly decrease with temperature to match experimental C-V trend, from $1.25 \times 10^{17}$ $cm^{-3}$ at 300 K to $9 \times 10^{16}$ $cm^{-3}$ at 77 K. c) Simulated electric field distribution in GeSn on Si APD structure at 77 K, at different reverse bias values. Inlet figure showed the electric field intensity in log scale. d), e) Avalanche gain as function of reverse bias extracted under 1.55 μm and 2 μm, respectively, from 77 K to 250 K.

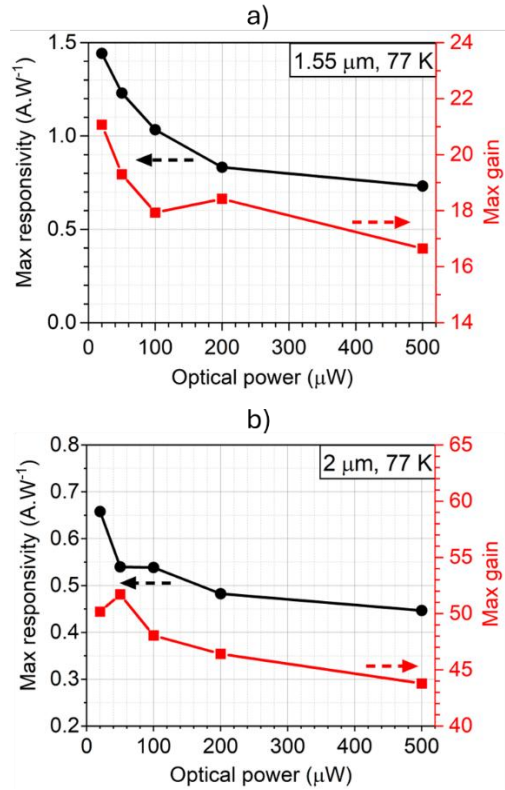

*Figure 4*. *Responsivity (black) and avalanche gain (red) as function of optical power at 77 K, under a) 1.55 μm illumination and b) 2 μm illumination*

**Discussion**

The significant improvement in detection range and 2 μm responsivity of GeSn on Si APD in this work was mainly attributed to the successful growth of relatively high Sn content GeSn absorber on top of thin Ge absorber. It can be considered as approaching the limit of GeSn direct epitaxial growth on Si [35,44,45]: amplified lattice mismatch effect between GeSn on Si via thin Ge buffer, compared to traditional GeSn growth on thick Ge buffer favored more intense plastic relaxation and SRE growth process, resulting in a significant increase of Sn content compared to the nominal target. Such approach can indeed be applied to grow even higher Sn content, as recently reported in Refs. [44,45] with record Sn content around 30%, pushing GeSn on Si SACM APD detection range towards MWIR range beyond 3 µm. However, it should be noted that the impact from threading dislocation also became more pronounced with higher Sn content. As shown in **Table 2**, the dark current density of GeSn on Si APD in this work remained higher than lattice-matched HgCdTe or Sb-based APD: while it can be partly attributed to the bigger size and absence of passivation of GeSn on Si APD device in this work, it was mainly due to strong trap-assisted tunneling (TAT) process originated from high threading dislocation density in both Ge buffer and GeSn absorber. In addition, the reader can notice low 1.55 µm and 2 µm primary responsivities despite a high absorption coefficient around 10000 cm$^{-1}$ (see ellipsometry data in **Supporting Information – Figure SI1**) and a relatively thick absorber of 250 nm: for example, 2 µm primary responsivities in this work were found between 0.011 A.W$^{-1}$ and 0.066 A.W$^{-1}$, well below theoretical value of 0.19 A.W$^{-1}$, estimated based on the absorption coefficient data, absorber thickness and surface reflection (R = 38%). It can be explained by partial depletion in the GeSn absorber due to high background p-doping concentration in GeSn absorber, as observed earlier in the simulated electric field distribution in **Figure 3c**. Background p-doping concentration was previously estimated to be around 10$^{17}$ cm$^{-3}$ for high Sn content GeSn, with common belief that it originated from defect-induced vacancy complexes and strongly related to the threading dislocation density in the GeSn layer [46]. Given high threading dislocation density observed from TEM images for GeSn grown on thin Ge buffer, both in this work and in our previous work **(Supporting Information – Figure SI5)** we expected that the background p-doping concentration in both GeSn and Ge layer of our APD structures is close to, if not even higher than 10$^{17}$ cm$^{-3}$, which eventually limited the device responsivity.

*Table 2. Comparison of dark current density $J_d$ (at avalanche gain M=10) between HgCdTe, Sb-based and GeSn on Si APD (this work). Details regarding device temperature and size were also included.*

| Material | $J_d$ @ M=10 (A.cm$^{-2}$) | Temperature (K) | Device size (μm) | Ref |
|---|---|---|---|---|
| HgCdTe | $1.0 \times 10^{-6}$ | 77 | ~ 10 | 3 |
|  | $3.0 \times 10^{-4}$ | 125 | 10 - 120 | 47 |
|  | $5.0 \times 10^{-3}$ | 80 | 64 | 8 |
| Sb-based | $5.0 \times 10^{-5}$ | 180 | 80 - 200 | 21 |
| **GeSn on Si (this work)** | $7.3 \times 10^{-2}$ (@ 1.55 μm) $2.6 \times 10^{-2}$ (@ 2 μm) | 77 | 350 |  |

Recent works reported a reduction of background p-doping concentration with thicker GeSn/Ge layer: based on C-V measurement data, Thai *et al.* reported a low background p-doping concentration between $10^{15}$ - $10^{16}$ for very thick GeSn absorber in photodiode structure (above 1 μm thick, grown on top of 1.2 μm thick Ge buffer) [48], while Tetzner *et al.* reported an background p-doping concentration close to, or even below $10^{15}$ cm$^{-3}$ in thick Ge buffer grown on Si (above 1 μm thick) [46]. Both results provided further proof on correlation between crystal quality – represented by threading dislocation density - and background p-doping concentration, with lower background p-doping concentration linked to a lower threading dislocation density in the upper GeSn/ Ge layer, distanced from SRE or half-loop dislocation region. Therefore, adjusting Ge buffer thickness is the key to improve GeSn on Si APD performance. A slightly thicker Ge buffer (300 nm - 500 nm) allowed growing thicker GeSn absorber, without severe trade-off in GeSn crystal quality. Thicker GeSn absorber with better crystal quality simultaneously increased light absorption volume and junction width, leading to higher responsivity. Further electric field drop in thicker Ge buffer can also be mitigated, if background p-doping concentration is reduced at the same time, similar to what was reported in Ref. [46]. In addition, based on our results, accidental diffusion of p+ doping from Si charge layer to Ge buffer did not inhibit APD operation. Therefore, in APD design with thicker Ge buffer, the first 100 nm of Ge layer can be used as the p+ charge layer instead of Si, as high density of half-loop dislocation increased the background p-doping concentration beyond $10^{17}$ cm$^{-3}$, which was the appropriate doping value for such layer. Finally, additional efforts will be focused on limiting the surface dark current: given the uncertainty regarding the efficiency of traditional passivating layer ($Al_2O_3$, $SiO_2$) on GeSn, e-SWIR transparent, thick top contact layer (i.e. high band-gap material, like Ge, Si or SiGeSn) can be utilized to increase the distance from the junction to the surface, thus reducing surface dark current while minimizing parasitic light absorption.

**Conclusion**

In this work, we demonstrated GeSn on Si SACM APD with detection range up to 2.7 μm, thanks to a GeSn absorber with high Sn content up to 12.7% Sn. The device showed high avalanche gains at 77 K up to 21 at 1.55 μm and up to 52 at 2 μm, alongside good responsivities up to 1.45 A.W$^{-1}$ at 1.55 μm and 0.66 A.W$^{-1}$ at 2 μm. Slight increase of Ge buffer thickness, with half-loop dislocation Ge region near Ge/Si interface utilized as p+ charge layer can be an efficient solution to reduce the background p-doping and to grow thick, high quality GeSn absorber, thus simultaneously reducing APD dark current and increasing its responsivity. These results confirmed the feasibility of high Sn content GeSn on Si APD for e-SWIR detection, with potential extension to MWIR range.

## METHODS

**Material growth and characterization**

The GeSn on Si SACM APD was epitaxially grown on a ASM Epsilon® 2000 RPCVD reactor, with $SiH_4$, $GeH_4$ and $SnCl_4$ as Si, Ge and Sn gas precursors. XRD 2θ-ω and (-2-2 4) RSM measurements were conducted using a Panalytical X'Pert Pro Materials Research Diffractometer. Ellipsometry measurement was conducted using a Variable Angle Spectroscopic Ellipsometer (WVASE32), ranging from 1300 to 2500 nm wavelength. PL measurement was conducted using a Horiba iHR320 spectrometer, equipped with an InSb detector.

**Device fabrication**

APD device was fabricated using standard photolithography to define the pattern, then the mesa structure was formed via wet etching using a mixture of HCl, $H_2O_2$ and deionized water. Etching automatically stops at Si charge layer. Ohmic contact at the top mesa surface and on the back side of n+ doped Si substrate, using 10/300 nm Cr/Au e-beam deposition.

**Device characterization**

APD characterization was performed from 77 K to 300 K, with I-V, C-V and spectral response measurement. I-V characteristics were measured with Keysight B2911B source measure unit (SMU). C-V measurements were conducted via a Keithley 590 CV Analyzer. For the calculation of APD gain, primary responsivities were taken at the reverse bias where capacitance drops and then stabilizes. 1.55 μm and 2 μm responsivities were measured with 1.55 μm laser from BKTel Photonics and continuous tunable laser from IPG Photonics. The incoming optical power was measured using Thorlabs S401C thermal sensor. Responsivities were calculated either from directly extracting dark I-V from light I-V, or from noise-rejection lock-in amplifier setup, with these results cross-checked against each other. Finally, spectral responses were measured using a ThermoFisher Nicolet 8700 FTIR, equipped with an IR source.

## AUTHOR INFORMATION

**Corresponding Author**

* Shui-Qing Yu

Email: syu@uark.edu

**Author Contributions**

S. -Q. Y, W. D., B. L., J. L. and C. -C. C. proposed and supervised the project. R. K., S. A. and C. -C. C. fabricated APD device. X. W. conducted simulation of APD device characteristics. P. C. G. and H. S. conducted TEM characterization and analysis. P. C. G. conducted XRD, ellipsometry and PL measurement of the sample. Q. M. T., J. R. and Y. Q. developed and implemented the device characterization setup. Q. M. T, J. R. and A. S. A. conducted the device characterization and analyzed the data. All authors discussed the results, commented on the manuscript, and gave approval to the final version of the manuscript.

**Competing interests**

The authors declare no competing interests.

**Funding Sources**

This work was supported by the Air Force Research Laboratory (AFRL)/AFWERX (Contract No. FA864922P0744), by United States Air Force (Contract No. FA865023C1140), and partially supported by Northrop Grumman Corporation (NGC) gift funding. J. Rudie was supported by DoW SMART Scholarship.

**Acknowledgement**

This work was supported by the Air Force Research Laboratory (AFRL)/AFWERX (Contract No. FA864922P0744) and United States Air Force (Contract No. FA865023C1140). Shui-Qing Yu would like to thank Northrop Grumman Corporation (NGC) for gift funding to partially support this work. He also gives his thanks for many fruitful discussions and encouragements from Drs. Alex Toulouse and Leye Aina at NGC for this work and greatly appreciates their input. J. Rudie acknowledges the funding support from the DoW SMART Scholarship. This work was performed, in part, at the Center for Integrated Nanotechnologies, an Office of Science User Facility operated for the U.S. Department of Energy (DOE) Office of Science. Los Alamos National Laboratory, an affirmative action equal opportunity employer, is managed by Triad National Security, LLC for the U.S. Department of Energy's NNSA, under contract 89233218CNA000001.

# Supporting Information

# Germanium-tin (GeSn) avalanche photodiode up to 12.7% Sn content for extended SWIR detection

*Quang Minh Thai [1], Rajesh Kumar [1], Justin Rudie [1,2], Xiaoxin Wang [3], Abdulla Said Ali [1,2], Perry C. Grant [4], Hryhorii Stanchu [5], Yunsheng Qiu [1], Steven Akwabli [1], Chun-Chieh Chang [6], Jifeng Liu [3], Baohua Li [4], Wei Du [1,2,5] and Shui-Qing Yu [1,2,5]\**

[1] Department of Electrical Engineering and Computer Science, University of Arkansas, Fayetteville, Arkansas 72701, USA

[2] Material Science and Engineering Program, University of Arkansas, Fayetteville, Arkansas 72701, USA

[3] Thayer School of Engineering, Dartmouth College, Hanover, New Hampshire 03755, USA

[4] Arktonics, LLC, 1339 S. Pinnacle Dr., Fayetteville, Arkansas 72701, USA

[5] Institute for Nanoscience and Engineering, University of Arkansas, Fayetteville, Arkansas 72701, USA

[6] Center for Integrated Nanotechnologies, Los Alamos National Laboratory, Los Alamos, New Mexico, 87545, USA

This supporting information contains data regarding photoluminescence (PL) and ellipsometry characterization of the APD sample (**Figure SI1**), responsivity as function of reverse bias under 1.55 µm and 2 µm illumination from 77 K to 300 K (**Figure SI2**), dark and light IV characteristics from 150 K to 300 K, under 1.55 µm illumination (**Figure SI3**) and under 2 µm illumination (**Figure SI4**), and a comparison of TEM image between this work and our previous APD structure [1] (**Figure SI5**).

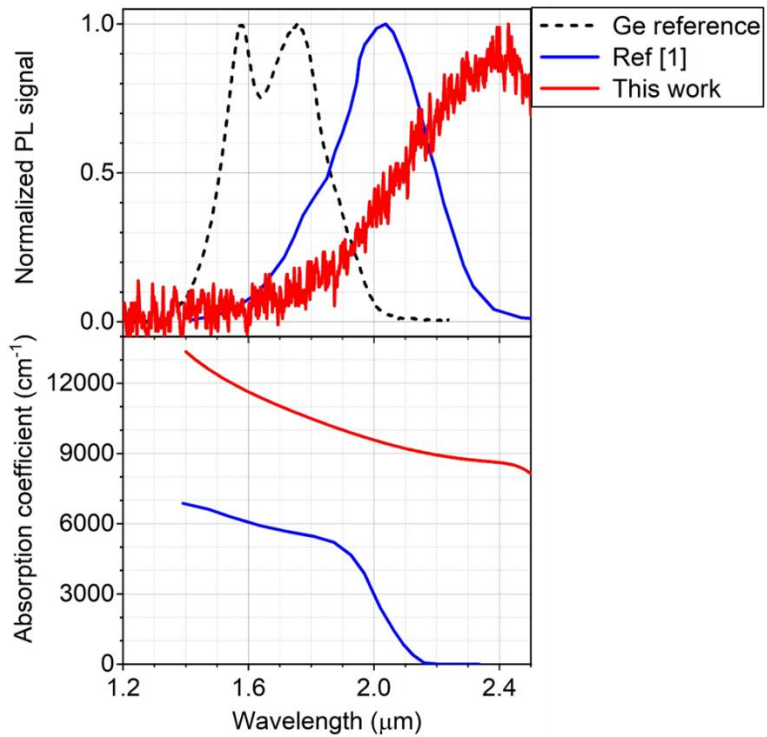

*Figure SI1*. <u>Top</u>: *PL spectrum (1064 nm optical excitation) of APD sample in this work (red, solid line), compared to a Ge reference (black, dashed line) and our previous work [1] (blue, solid line).* <u>Bottom</u>: *Absorption coefficient extracted from ellipsometry characterization of APD sample in this work (red), compared to our previous work [1] (blue). Please note that the PL and absorption coefficient spectrum range stops at 2.5 µm due to instrument limitation.*

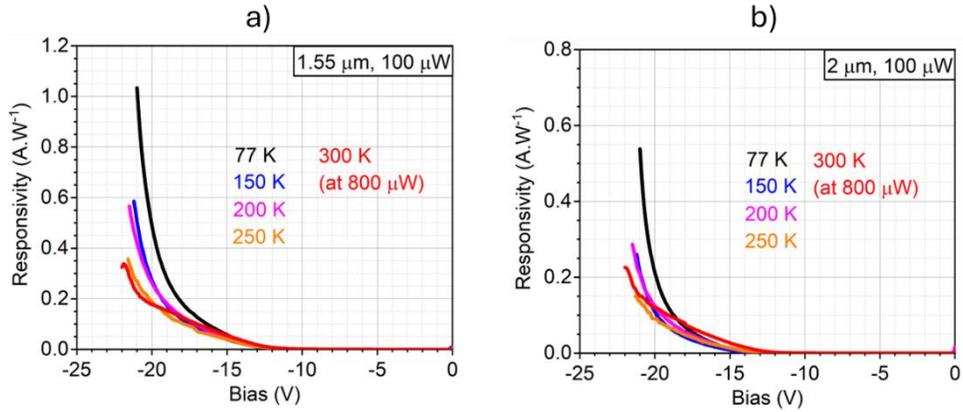

*Figure SI2*. a) 1.55 μm responsivity and b) 2 μm responsivity as function of reverse bias, respectively, from 77 K to 300 K. 100 μW optical power is used to characterize responsivity from 77 K to 250 K, while 800 μW optical power is used to characterize responsivity at 300 to provide better distinction between dark and light IV (see Figures SI3,SI4 below)

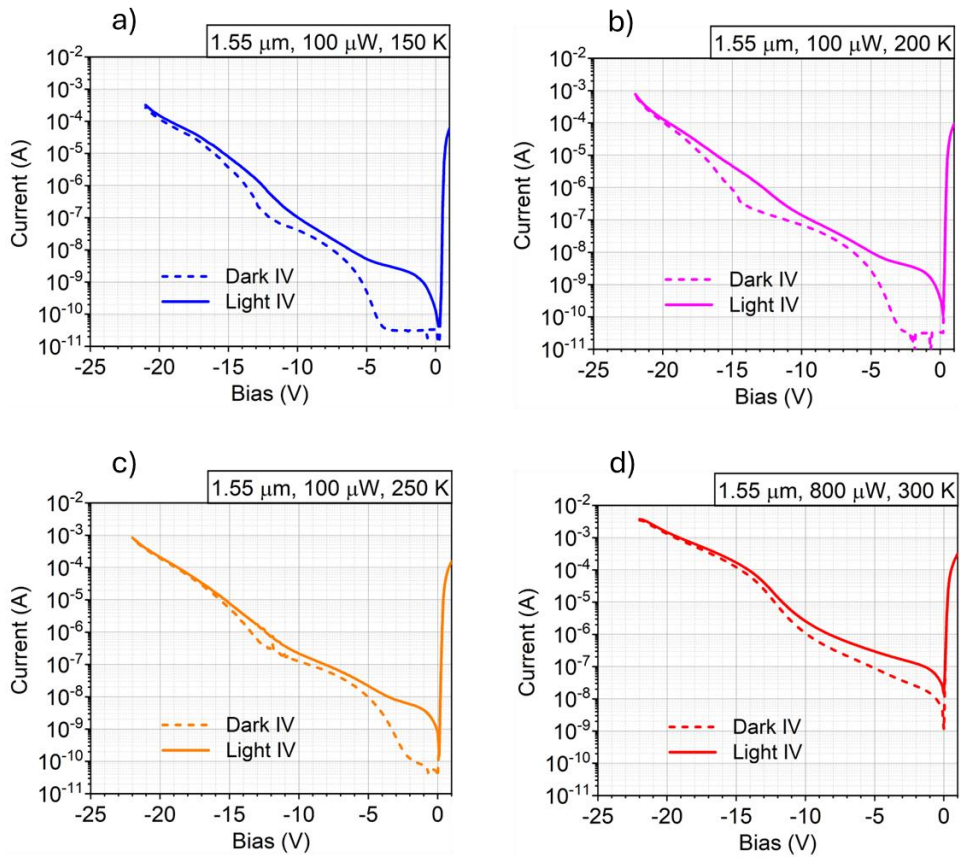

*Figure SI3*. Dark (dashed lines) and light IV (solid lines) characteristics under 1.55 μm illumination from 150 K to 300 K. 100 μW of optical power was used for characterization from 150 K to 250 K, while 800 μW of optical power was used for characterization at 300 K.

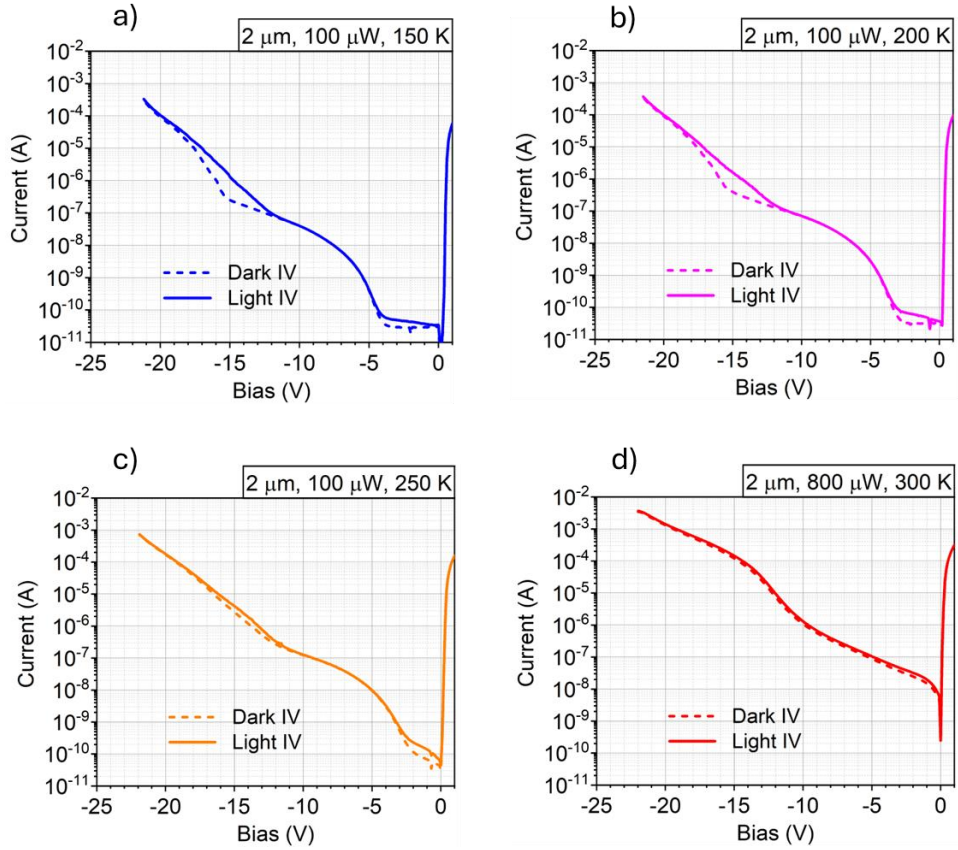

*Figure SI4*. Dark (dashed lines) and light IV (solid lines) characteristics under 2 μm illumination from 150 K to 300 K. 100 μW of optical power was used for characterization from 150 K to 250 K, while 800 μW of optical power was used for characterization at 300 K.

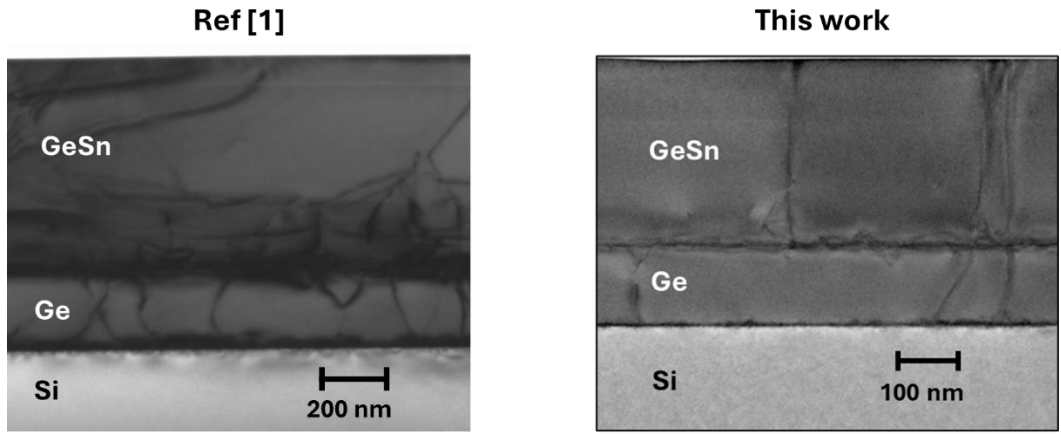

*Figure SI5*. Comparison of TEM images of APD structures from our previous work [1] (left) and this work (right). For both Ge buffer thickness (241 nm and 122 nm), threading dislocation propagating through the buffer can be observed. On the GeSn absorber, half loop dislocation can be observed near the Ge/GeSn interface in our previous work, possible due to a thicker absorber (518 nm compared to 250 nm in this work).